# Effects of Poly(styrene/pentafluorostyrene-block-vinylpyrrolidone) Amphiphilic Kinetic Hydrate Inhibitors on the Dynamic Viscosity of Methane Hydrate Systems at High-Pressure Driving Forces

*Chong Yang Du, André Guerra, Adam McElligott, Milan Marić, Phillip Servio\**

Department of Chemical Engineering, McGill University, Montreal, Quebec H3A 0C5, Canada

*phillip.servio@mcgill.ca

## Abstract

Reversible addition-fragmentation chain-transfer polymerization with a switchable chain-transfer agent was employed to synthesize amphiphilic block copolymers poly(styrene-b-vinylpyrrolidone) and poly(pentafluorostyrene-b-vinylpyrrolidone) at 10 wt.% hydrophobic content as kinetic hydrate inhibitors for methane hydrates. The dynamic viscosity of methane hydrate slurries was measured in a high-pressure rheometer up to 15 MPag. At 700 ppm of additives in aqueous media, the relative time for slurries to grow to 200 mPa·s was 2.2-2.4 times longer than water reference values for the block copolymers. In contrast, it was only 1.3 for the poly(vinylpyrrolidone) homopolymer, demonstrating a reduced tendency for hydrate particle adhesion in block copolymer solutions. By increasing the concentration to 7000 ppm, however, the relative time only increased to 2.6-2.7. On the other hand, a block copolymer with 5 wt.% poly(pentafluorostyrene) block at 7000 ppm reached 3.5, which may indicate that the optimal hydrophobic content differs for each amphiphilic polymer and depends on monomer selection. No significant effect of polymer aggregation on hydrate growth was observed for the copolymers used in this study, which had the same hydrophobic percentage and molecular weights between 10,000 and 40,000 g/mol.

# 1. Introduction

Gas hydrates are crystalline clathrate compounds formed due to hydrogen bonding between water molecules, which organize into matrices at high pressures and low temperatures. By trapping small gas molecules such as methane, ethane and propane inside their cavities, water's cage-like structures can be stabilized via van der Waals forces. Many components of natural gas, combined with the accumulation of free water in pipelines and the operating conditions in offshore facilities, are favourable for gas hydrates formation and lead to pipeline blockages.[1-3]

For flow assurance purposes, research on hydrate inhibition is important to prevent large scale production losses, equipment damage, and safety hazards.[2] Traditional antifreezes such as methanol and ethylene glycol are used as thermodynamic hydrate inhibitors in industry. They shift the thermal equilibrium for phase transition to lower temperatures and higher pressures than those in pipelines but are required in large quantities, typically between 20-50 wt.%.[4,5] Studies on low dosage hydrate inhibitors (LDHIs), effective at 0.01-5 wt.%, are therefore crucial to reduce transportation and treatment costs, as well as the ensuing environmental impacts.[5]

LDHIs are generally divided into two broad categories: kinetic hydrate inhibitors (KHIs) and anti-agglomerants (AAs).[4] KHIs are typically water-soluble polymers, such as poly(vinylpyrrolidone) (PVP) and poly(vinylcaprolactam) (PVCap), two structurally similar polymers that are commercially available and widely researched. The amide groups allow PVP to engage in hydrogen bonding, an ability that perturbs local water structure and, in turn, extends hydrate induction time.[4,6] Furthermore, computer simulations have modelled PVP's ability to adsorb onto hydrate surfaces, and to sterically hinder the subsequent incorporation of water and gas molecules into the hydrate phase.[6,7] Experimentally, significant reduction in methane consumption rates has been observed with PVP/PVCap-inhibited systems compared to pure water by adding less than 1 wt.% of KHIs to the solution.[8-10]

The lower critical solution temperature of PVP occurs above water's boiling temperature, hence the cloud point will not be reached and polymer aggregation will not take place at the cold temperatures of pipeline operating conditions.[11, 12] Nevertheless, rheological studies on the viscosity of hydrate slurries in aqueous systems in the presence of PVP/PVCap have shown signs of sudden agglomeration with increasing additive concentration, despite the decrease in methane consumption compared to the control system without additives.[13, 14] The viscosity of hydrate slurries has been perceived as one of the main indications of the degree of risk for pipeline blockage, thus the ability of reducing particle clustering is crucial for hydrate inhibitors.[15] In recent years, a novel type of hydrate inhibitor has been proposed, which contains a short hydrophobic segment added to a hydrophilic polymers. This provides the KHIs an amphiphilic structure analogous to that of surfactants, which makes them capable of maintaining a dispersed, low-viscosity slurry during transportation, similar to the effects of AAs.[4, 16] The involvement of both hydrophobic and hydrophilic functionalities is beneficial for the interactions and disruption of water and hydrate structures.[17, 18] Moreover, copolymers with greater hydrophobicity, that also ensure a high enough cloud point for the application, often exhibit better inhibition performance in terms of lower onset temperature at which hydrate formation trigger.[19, 20]

Controlled polymerization using chain-transfer agents (CTAs), notably via reversible addition-fragmentation chain-transfer (RAFT) polymerization, to synthesize amphiphilic block copolymers with hydrophobic segments such as poly(styrene) (PS), poly(pentafluorostyrene) (PPFS) and poly(ε-caprolactone) and water-soluble segments such as PVP, PVCap and poly(vinyl alcohol) have been achieved in literature.[9, 10, 16, 21] At 1-20 wt.% hydrophobic content, these amphiphilic KHIs demonstrated improved anti-nucleation properties and further decreased methane consumption during the growth phase compared to hydrophilic homopolymers with an increased hydrophobic content in the copolymer.

In the scope of this study, PS/PPFS-b-PVP block copolymers were synthesized via RAFT polymerization and tested as a new class of KHIs in a high-pressure rheometer. Methane hydrate growth was recorded and measured in terms of dynamic viscosity with respect to time. The effectiveness of the amphiphilic KHIs to maintain a low-viscosity hydrate slurry and to delay particle clustering was investigated and compared with the performance of a PVP homopolymer examined in a previous work. The main goal of this work is to examine whether these amphiphilic polymer additives can combine the anti-nucleation property of KHIs with the anti-agglomeration property of surfactants.

## 2. Materials and Methods

2.1 Materials

This work used a variety of chemicals and gases which are listed here with their abbreviations, sources, purities, and other identification information: N-Vinyl-2-pyrrolidone (NVP monomer, Acros Organics, 99%, stabilized with NaOH), Styrene (St monomer, Sigma-Aldrich, ReagentPlus®, ≥99%, contains 4-tert-butylcatechol as stabilizer), 2,3,4,5,6-Pentafluorostyrene (PFS monomer, Sigma-Aldrich, 99%, contains 0.1% *p*-tert-butylcatechol as inhibitor), 2-Cyanopropan-2-yl N-methyl-N-(pyridine-4-yl)carbamodithioate (Switchable RAFT agent or CTA, Sigma-Aldrich, high-performance liquid chromatography (HPLC) grade, 97%), Azobisisobutylnitrile (AIBN, recrystallized), Trifluoromethanesulfonic acid (Triflic acid, MilliporeSigma, ReagentPlus®, ≥99%, in ampule), 4-Dimethylaminopyridine (Tokyo Chemical

Industry, >99.0%), Acetonitrile (MeCN, Fisher Scientific, Certified ACS), Tetrahydrofuran (THF, Sigma-Aldrich, HPLC grade, ≥99.9%), Methanol (Fisher Chemical, HPLC grade), Hexanes (Fisher Chemical, Certified ACS), Reverse osmosis water (RO water, final conductivity of 10 µS, maximum organic content of 10 ppb), Nitrogen (ALPHAGAZ™, Smart Top), Methane (MEGS Specialty Gases and Equipment Inc., ultra high purity, 99.99%).

2.2 Synthesis of amphiphilic block copolymers

The amphiphilic block copolymers were synthesized using the switchable RAFT agent 2-cyanopropan-2-yl N-methyl-N-(pyridine-4-yl)carbamodithioate, which can polymerize more activated monomers (MAMs) such as styrene and PFS in its protonated form, and can chain extend less activated monomers (LAMs) like NVP and N-vinylcaprolactam (NVCap) when deprotonated.[22] The reaction scheme is illustrated in Figure 1, and a strong acid such as triflic acid is required to perform the "switch".

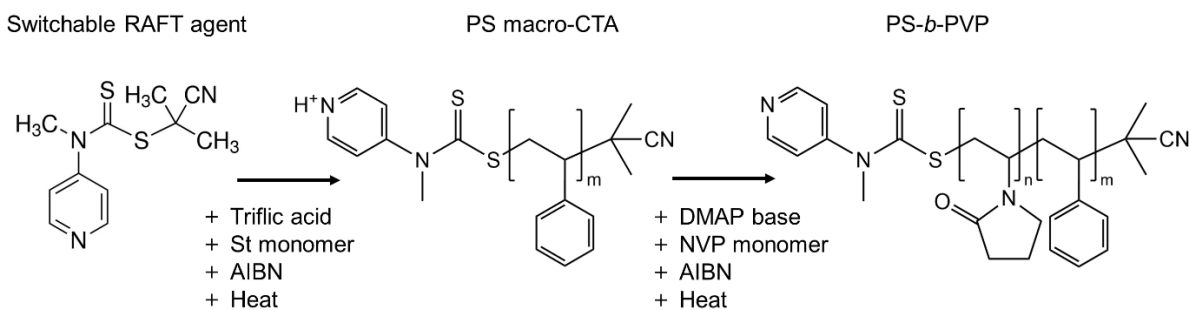

Figure 1 Reaction scheme for synthesis of poly(styrene-b-vinylpyrrolidone) using switchable RAFT agent.

Similar to the methods developed by Benaglia et al. (2009),[22] a stock solution of triflic acid in MeCN was prepared by transferring 115.0 g of MeCN in a glass round-bottom flask, submerging the flask in an ice water bath, and purging with nitrogen for 30 minutes, then quickly pouring 5 g of triflic acid into the flask right after breaking open the ampule. The solution was purged with nitrogen for another 30 minutes, reweighed to obtain the exact acid concentration (4.7 g), and sealed with parafilm. This step prevents the triflic acid from reacting violently with moisture in the air once the container is unsealed and allows heat dissipation during protonation of the RAFT agent. Macro-chain transfer agents (mCTAs) of PS and PPFS were synthesized using the conditions presented in Table 1, some of which are similar to those in Rajput et al. (2018), which also reported the critical micelle concentrations (CMCs) for the block copolymers synthesized using a similar methodology.[9] The number following the polymer abbreviation refers to their molecular weight (MW) in kg/mol. The target MW is defined as monomer to CTA ratio multiplied by the estimated conversion at the end of the reaction. The MAM mCTAs were dissolved using THF and precipitated using methanol.

Table 1 Reagent composition and reaction conditions for synthesis of mCTA of PS and PPFS.

| Polymer ID | Reagents (molar equivalent) | | | | $T$ ($\pm 5$ °C) | Time (min) | Target MW (g/mol) |
|---|---|---|---|---|---|---|---|
| | Monomer | RAFT agent | Triflic acid | AIBN | | | |
| PS4 mCTA | 100 | 1 | 1 | 0.2 | 90 | 240 | 4,000 |
| PPFS4 mCTA | 50 | 1 | 1 | 0.2 | 90 | 240 | 4,000 |
| PPFS2 mCTA | 50 | 1 | 1 | 0.1 | 90 | 60 | 2,000 |
| PPFS1 mCTA | 12.5 | 1 | 1 | 0.1 | 90 | 180 | 1,000 |

After a purification process by dissolution, reprecipitation and decantation, the mCTAs were left in the fume hood overnight and dried in vacuum oven for 48 hours at 50 °C. They were then deprotonated using a stoichiometric amount of DMAP base and chain extended with NVP monomer in bulk polymerization using the conditions in Table 2. The amphiphilic block copolymers were dissolved in THF and precipitated using hexanes.

Table 2 Reagent composition and reaction conditions for synthesis of PS/PPFS-b-PVP block copolymers.

| Polymer ID | Reagents (molar equivalent) | | | | $T$ (±5 °C) | Time (min) | Target total MW (g/mol) |
| --- | --- | --- | --- | --- | --- | --- | --- |
| | Monomer | mCTA | DMAP base | AIBN | | | |
| PS4-b-PVP40 | 1025 | 1 | 1 | 1 | 75 | 90 | 40,000 |
| PPFS4-b-PVP40 | 1150 | 1 | 1 | 1 | 75 | 100 | 40,000 |
| PPFS2-b-PVP20 | 580 | 1 | 1 | 0.5 | 70 | 180 | 20,000 |
| PPFS1-b-PVP10 | 310 | 1 | 1 | 0.3 | 70 | 240 | 10,000 |

In terms of mass ratio, typically 30.0 g of NVP monomer was added to 1.0 g of mCTA, with a target conversion of 30% and a target percentage of 10 wt.% for the hydrophobic segment in the copolymer. Above this conversion, the reaction mixture usually becomes too viscous for the magnetic stir bar to ensure proper heat and mass transfer, and thermal runaway might occur when conversion approaches 40%. Solution polymerization was attempted to decrease viscosity thus increase final conversion. However, while the addition of 50 vol.% of MeCN successfully allowed conversion to reach over 65% for RAFT polymerization of PVP homopolymer, the reaction mixture with MeCN using PS mCTA4 reached a plateau of around 10% conversion after 10 hours and did not improve until a total of 20 hours of reaction with NVP chain extension. For the purposes of our study and the availability of mCTAs, we did not pursue using other solvents such as toluene, which are suitable for PVCap homopolymer synthesis.[23]

Polymerization reactions were conducted in Dima Glass 24/40 three-neck round-bottom flask reactor, with Dima Glass water-jacketed condenser cooled to 1.00 °C attached to the middle neck. Reaction mixtures were nitrogen purged for 30 minutes before increasing the temperature using a Glas-Col lower hemispherical cloth heating mantle and temperature controller. The nitrogen purge needle and thermocouple were inserted in the reactor through rubber septa which sealed the right and left necks. The reaction mixture was stirred using a magnetic stir bar at 500 rpm using an Isotemp stir plate by Fisher Scientific.

2.3 Characterization

The conversions were calculated using proton nuclear magnetic resonance ($^1$H NMR) with a Bruker Avance III HD 500 MHz NMR Spectrometer by integrating the polymer peaks with respect to vinyl proton peaks of the monomer. The selected deuterated solvent was chloroform-d (CDCl$_3$).

Molecular weights and dispersity (Đ) of MAM mCTAs were characterized using gel permeation chromatography (GPC) with THF at 40 °C as eluent and relative to PS standards for calibration. Samples were prepared at 2 mg/mL in 1 dram vials. The GPC equipment is composed of Waters 717plus Autosampler, Waters 1515 Isocratic HPLC Pump, Waters 2487 Dual λ Absorbance Detector, and Waters 2414 Refractive Index Detector.

Molar composition of PPFS-b-PVP copolymers were determined using quantitative NMR analysis using external standards with known concentrations. The concentration of fluorine in the PPFS block in a sample was calculated relative to the Bruker standard 0.1 M of CF$_3$CO$_2$H in CDCl$_3$ using $^{19}$F quantitative NMR. The concentration of protons in the PVP block of the same sample was calculated relative to the Bruker standard $^{31}$P with 0.0485 M TPP in Acetone-d6. The composition of PS-b-PVP copolymer was determined by integrating their respective proton broads in $^1$H NMR ($\delta$, CDCl$_3$): 7.26-6.86 (3H, C$_6$*H*$_5$CHCH$_2$, phenyl group in PS block), 6.86-6.25 (2H, C$_6$*H*$_5$CHCH$_2$, phenyl group in PS block), 4.11-3.48 (1H, C$_4$H$_6$NOC*H*CH$_2$, backbone in PVP block), 3.48-2.92 (2H, C$_4$*H*$_6$NOCHCH$_2$, pyrrolidone group in PVP block). The weight composition and molecular weight were then estimated based on the calculated molar composition and NVP conversion. Particle size of the block copolymers were determined via dynamic light scattering (DLS) using Zetasizer Nano Series (Nano-ZS) with a 4 mW 632.8 nm red laser. Three measurements were performed with 13 runs each to obtain the average particle size.

2.4 Viscosity measurements of methane hydrate slurries

Aqueous solutions of 700 and 7000 ppm by weight were prepared for the block copolymers by mixing for 24 hours at 500 rpm using a magnetic stirrer. The viscosities of hydrate-forming systems were measured using an Anton Paar Modular Compact Rheometer (MCR) 302 with a double-gap (DG) geometry and a pressure rating of 40 MPa. A 7.5 mL test sample was transferred into the cup-and-bob geometry using a 10 mL syringe and brought to the setpoint temperature using a Julabo F32 chiller with 50/50 volume mixture of ethylene glycol and water. Air was removed from the system by purging with methane at 0.5 MPag five times for 30 seconds. The system pressure was maintained by a Schlumberger DBR positive displacement pump. The hydrate growth experiment setup is illustrated in Figure 2.

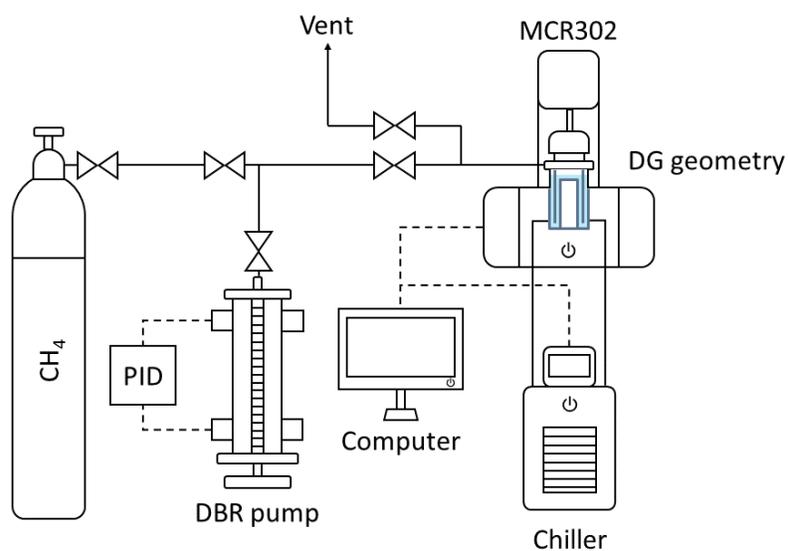

Figure 2 Experimental setup for methane hydrate growth and viscosity measurement.

The conditions selected for hydrate formation were temperatures of 0, 2, 4, 6 °C and pressures of 10, 15 MPag. The shear rate was chosen to be 400 s$^{-1}$ to ensure a short induction time,[24] which typically occurred within minutes, and to ensure high accuracy readings of the low-viscosity aqueous solutions. A safety torque limit of 115 mN·m was imposed on the measuring system to protect the geometry from the sudden viscosity increase due to phase change. When hydrate slurries reached about 1000 mPa·s, the measurements ceased. Temperature, pressure, and viscosity as functions of time were monitored and recorded using the Anton Paar RheoCompass software, which output csv data to be analyzed with MATLAB. All test conditions are summarized in Table 3 and were selected to allow direct comparison with data from previous studies.[14, 25] Three runs per condition were performed to verify reproducibility, and each hydrate growth experiment was allowed a 90-minute timespan.

Table 3 Test conditions for rheological experiments with hydrate forming systems.

| Concentration (ppm) | 700, 7000 |
|---|---|
| $T$ (±0.25 °C) | 0, 2, 4, 6 |
| $P$ (±0.2 MPag) | 10, 15 |
| Shear rate (s$^{-1}$) | 400 |
| Runs per condition | 3 |

## 3. Results and Discussion

3.1 Synthesis of amphiphilic block copolymers

The conversion, number average molecular weight, as well as dispersity of PS and PPFS mCTAs are presented in Table 4. The estimated molecular weights using the conversion from $^1$H

NMR ($M_{n,\text{NMR}}$) agree with the number average molecular weights calculated from PS standards in THF GPC ($\bar{M}_{n,\text{GPC}}$), providing confidence in the estimation of $M_n$ of the block copolymers later on using NMR. The linearized conversion trendlines from Figure 3 indicate that first order kinetics can describe the RAFT polymerization for mCTAs and chain extension reactions at low conversions. In both cases, however, deviation from linearity was observed at higher conversions, potentially due to irreversible chain termination and to the greater viscosity, inducing less efficient heat transfer, respectively.

Table 4 Conversion, molecular weight, and dispersity of PS and PPFS mCTAs.

| Polymer ID | Conversion (%) | $M_{n,\text{NMR}}$ (g/mol) | $\bar{M}_{n,\text{GPC}}$ (g/mol) | Đ |
|---|---|---|---|---|
| PS4-b-PVP40 | 36.0 | 3,752 | 3,800 | 1.15 |
| PPFS4-b-PVP40 | 45.9 | 4,459 | 4,200 | 1.41 |
| PPFS2-b-PVP20 | 22.4 | 2,174 | 2,200 | 1.55 |
| PPFS1-b-PVP10 | 45.1 | 1,094 | 1,200 | 1.20 |

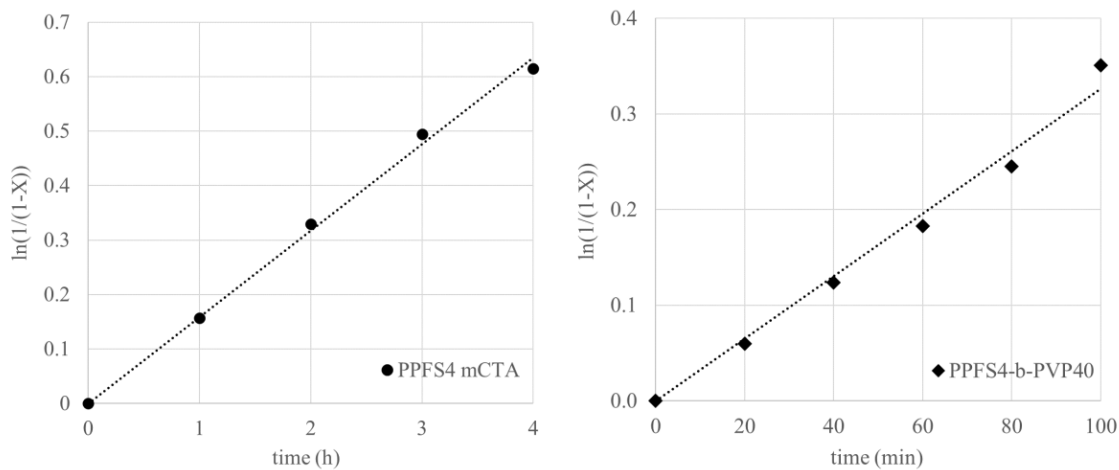

Figure 3 Plot of linearized conversion for PPFS4 mCTA (left) and PPFS4-b-PVP40 (right).

RAFT polymerization allows chain quality control and can achieve narrow molecular weight distributions (Đ < 1.5).[26] This applies to the mCTAs with the exception of PPFS2. The variation in Đ is likely related to the quality of the triflic acid stock solution, which quickly turned from colorless to yellow after only 2 weeks. After first synthesizing PS4 and then PPFS4, the difference in Đ (1.15 in comparison to 1.41) despite the similar $\bar{M}_n$ was assumed to be caused by the molar ratio of monomer to AIBN, which was 100:0.2 and 50:0.2 respectively. However, after preparing a new batch of triflic acid stock solution and synthesizing first PPFS1 then PPFS2, there was no correlation between Đ and initiator concentration, since Đ = 1.20 was obtained for a PPFS:AIBN ratio of 12.5:0.1 and Đ = 1.55 for a ratio of 50:0.1. It seems to be rather dependent on the age of the acid and thus its ability to protonate the RAFT agent simultaneously.

The conversions and molar compositions of the chain extended block copolymers were obtained from NMR spectra, then the total molecular weights and weight compositions were calculated based on these data. All values are summarized in Table 5. The target conversion of 30% for 30.0 g of NVP monomer reacting with 1.0 g of mCTA resulted in approximately 10 wt.% of hydrophobic block composition, confirmed with quantitative NMR analysis for PPFS-b-PVP and $^1$H NMR integration for PS-b-PVP as shown in Figure 4.

Table 5 Conversion, estimated molecular weight and composition of block copolymers.

| Polymer ID | NVP conversion (%) | $M_{n,NMR}$ (kg/mol) | mol% of mCTA | wt.% of mCTA |
|---|---|---|---|---|
| PS4-b-PVP40 | 29.8 | 37.7 | 9.4 | 8.9 |
| PPFS4-b-PVP40 | 29.6 | 41.9 | 6.1 | 10.1 |
| PPFS2-b-PVP20 | 28.6 | 20.7 | 5.9 | 9.8 |
| PPFS1-b-PVP10 | 24.2 | 9.8 | 5.5 | 9.2 |

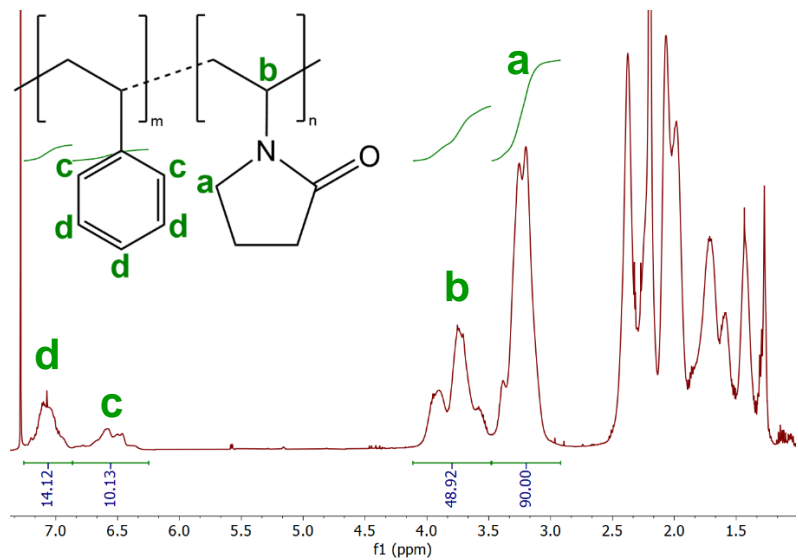

Figure 4 $^1$H NMR analysis for the molar composition of PS4-b-PVP40 block copolymer in CDCl$_3$

3.2 Viscosity measurements of methane hydrate slurries

    The dynamic viscosities of methane hydrates formed from aqueous solutions of 700 ppm of PS/PPFS4-b-PVP40 are compiled in Figure 5. The profiles shown in Figure 5 were all selected as the median run within the 3 repeated runs to reach agglomeration. In comparison to previous reports using water[25] and PVP40[14], an overall longer growth period was observed with these amphiphilic block copolymers.

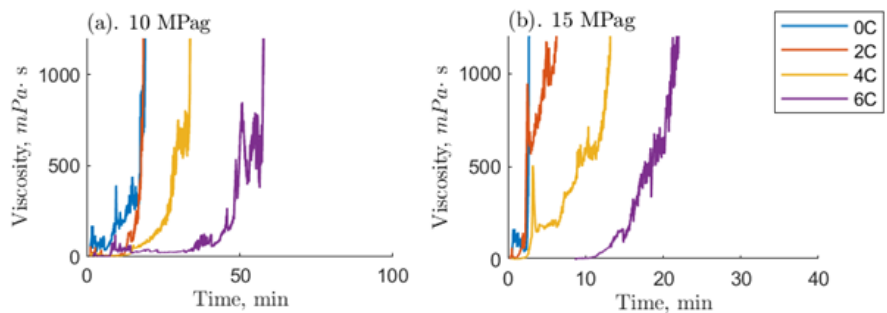
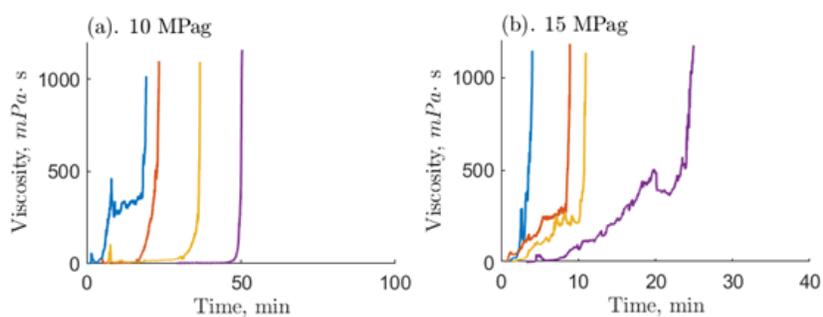
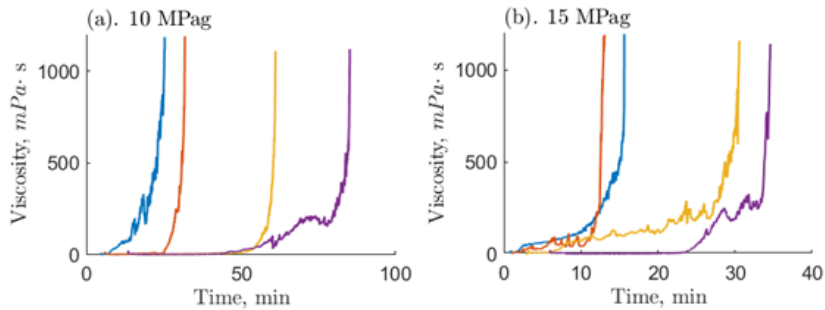
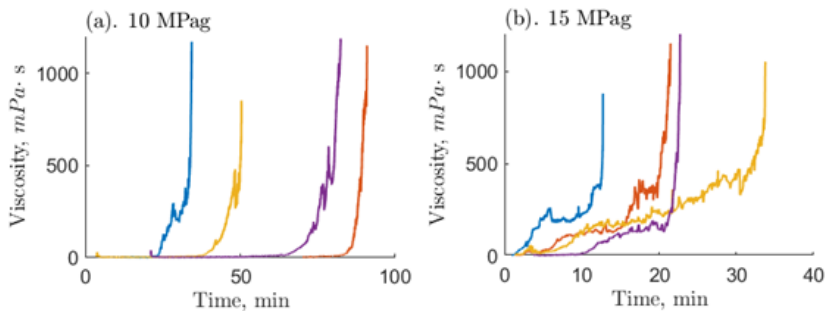

Figure 5 Viscosity profiles of methane hydrate slurries with respect to time for 4 test solutions: 1. Pure water (data from Guerra et al. 2022[25]), 2. 700 ppm PVP40 (data from Du et al. 2022[14]), 3. 700 ppm PS4-b-PVP40 and 4. 700 ppm PPFS-b-PVP40.

Previous studies have defined the dynamic viscosity profiles as typically comprising of a slurry phase followed by an agglomeration phase. Although particle size was not directly measured, it was assumed that the sharp viscosity rise at the end of the curve was due to sudden particle clustering, since the methane consumption rate is expected to be linear.[9, 24] The large aggregates can likely be attributed to capillary action of water molecules surrounding the hydrate particles.[15] With the novel amphiphilic KHIs, the same behaviour of sudden agglomeration persisted. Nevertheless, it is evident that adding a hydrophobic block extended the slurry phase to longer times, reaching near-solidification in Figure 5 and 200 mPa·s in Figure 6.

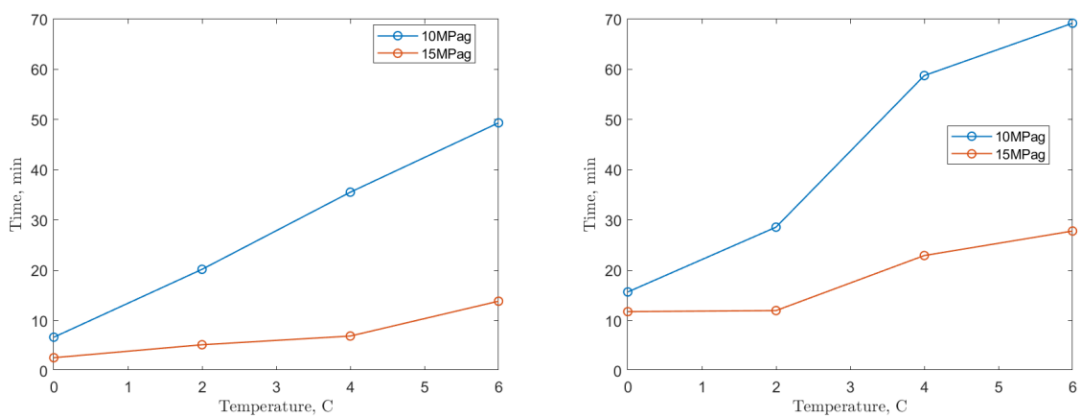

Figure 6 Median time required for methane hydrate slurries to reach 200 mPa.s for Left: PVP40 700 ppm (data from Du et al., 2022[14]) and Right: PS4-b-PVP40 700 ppm.

The time for the hydrate slurry to reach 200 mPa·s in each run was compiled, and related to a water reference. The dimensionless variable of relative time is used to compare the effectiveness of different additives. The relative time $t_{200}$ is defined as a ratio of the time required for an inhibited system to reach 200 mPa·s over the time required by pure water. This viscosity was selected for the water system as a representative value estimated to be the turning point from the slurry phase to the agglomeration phase.[25]

The $t_{200}$ values were calculated for all test conditions including all three runs. A total of 24 runs per polymer were compiled in Figure 7. The difference between the homopolymer and the copolymers is significant. The geometric mean of $t_{200}$ for the PVP40 700 ppm set was 1.3, and it increased to 2.4 and 2.2 for the PS4-/PPFS4-b-PVP40 sets, respectively, achieving a nearly 100% increase in growth time. Therefore, the addition of a PS block had a similar effect to adding a PPFS block at 700 ppm.

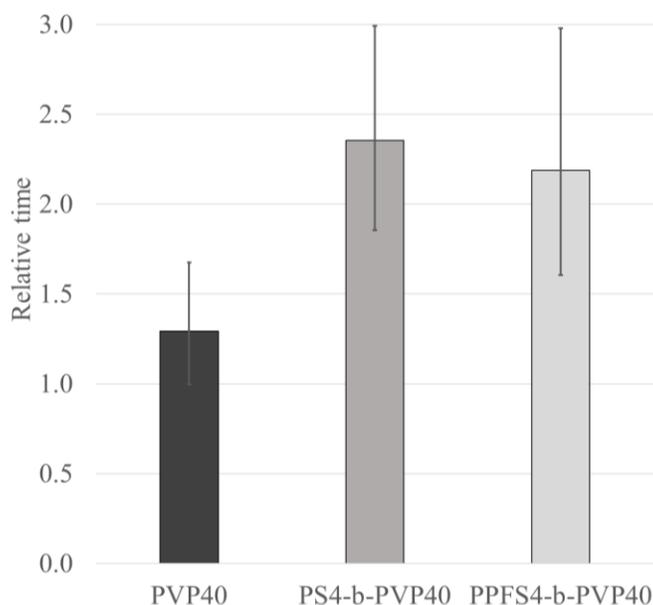

Figure 7 Geometric mean of relative time for methane hydrate slurries to reach to 200 mPa·s for PVP40 (data from Du et al. 2022[14]), PS4-b-PVP40 and PPFS4-b-PVP40 at 700 ppm, with 95% confidence interval error bars.

In literature, the addition of a more hydrophobic block copolymer (i.e., in terms of the selection of monomer and the percentage of hydrophobic block in the copolymer) resulted in a noticeably longer nucleation time and lower methane consumption rate compared to pure water and PVP/PVCap homopolymers.[10, 16] Thus, it was expected that the copolymer with a PPFS segment, being more hydrophobic than PS, would be more efficient in delaying hydrate agglomeration. Contrary to these expectations, Figure 7 presents comparable compiled results for PPFS and PS mCTAs.

PPFS4-b-PVP40 at 700 ppm did exhibit a phenomenon that was not observed with PS4-b-PVP40, nor with PVP40. Within a total of 24 hydrate growth runs per polymer, on 3 occasions the hydrate particles seemingly dissociated after nucleation with addition of PPFS4-b-PVP40, as presented in Figure 8. It occurred at 2 °C and 15 MPag, at 2 °C and 10 MPag, and at 4 °C and 10 MPag. These anomalous cases were not compiled in Figure 7 as the viscosity never reached 200 mPa·s despite the presence of a nucleation event.

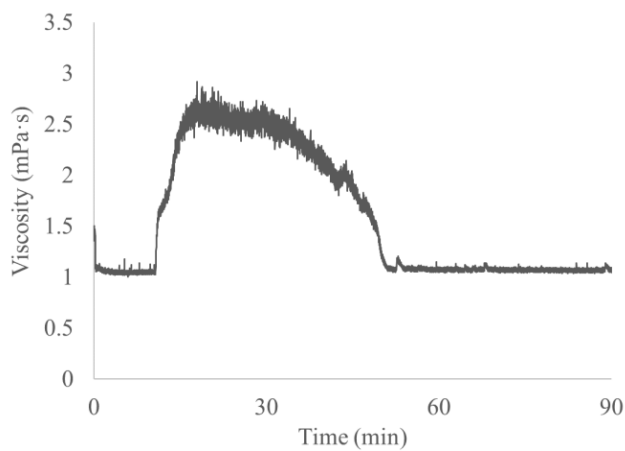

Figure 8 Signs of nuclei decomposition observed in methane hydrate growth experiments with addition of 700 ppm PPFS4-b-PVP40, figuring the case at 2 °C 15 MPag.

These instances of atypical yet desirable behavior were possibly due to the randomness induced by the turbulence in the test samples that occurs at the high shear rate of 400 s$^{-1}$. The amphiphilic KHIs either successfully deformed the structure of the nuclei or shielded them as they appeared, and prevented them from acquiring additional water and methane molecules to carry on their growth. Therefore, the maximum viscosity observed remained very low (<10 mPa·s) throughout the entire experimental timespan of 90 minutes. This behavior of nuclei dissociation can also be used to explain the outlier at 2 °C and 10 MPag with PPFS4-b-PVP40 in Figure 5. The reason why its viscosity profile lasted longer than the lowest subcooling condition (6 °C, 10 MPag) may be that after nucleation and the first viscosity rise, its viscosity slowly fell back to around 1 mPa·s, then potentially "re-nucleated" or resumed growth a second time to eventually reach agglomeration.

Additionally, no nucleation (i.e., change in viscosity) was observed on 3 other occasions for PPFS4-b-PVP40. This occurred once at 4 °C and 15 MPag and twice at 6 °C and 10 MPag. In other words, hydrate growth with PPFS4-b-PVP40 was observed for only 18 of the 24 cases (75% of the time), but in all cases for its PS counterparts, hence the slightly larger error bars and lower inhibitory performance in Figure 7. Despite these peculiarities, since their occurrence was not under control and attributed to a degree of randomness, in the scope of this study, PPFS or PS mCTA end groups behaved similarly.

Experiments were then performed to verify whether these phenomena of no growth and no nucleation would occur more frequently if the concentration of KHIs was increased from 700 ppm to 7000 ppm. The hypothesis was disproven, and the inhibition effect did not improve significantly despite a greater abundance of KHIs in the system. The number of cases of no growth/no nucleation was ≤6 and the geometric mean of $t_{200}$ was 2.6-2.7 for all three 10 wt.% PPFS-b-PVP copolymers at different molecular weights, as shown in Figure 9, which is very similar to that at 700 ppm. The high hydrophobic content in water may have hindered the anti-nucleation ability of PVP. Despite the stochastic nature of induction, commercial PVP40 at 7000 ppm only successfully nucleated 11 times out of the 24 90-minute experiments, which was more than twice as effective as the block copolymers at the same concentration.[14] Due to the limited number of successful formation events, which resulted in very large error bars, data for PVP40 is not presented in Figure 9.

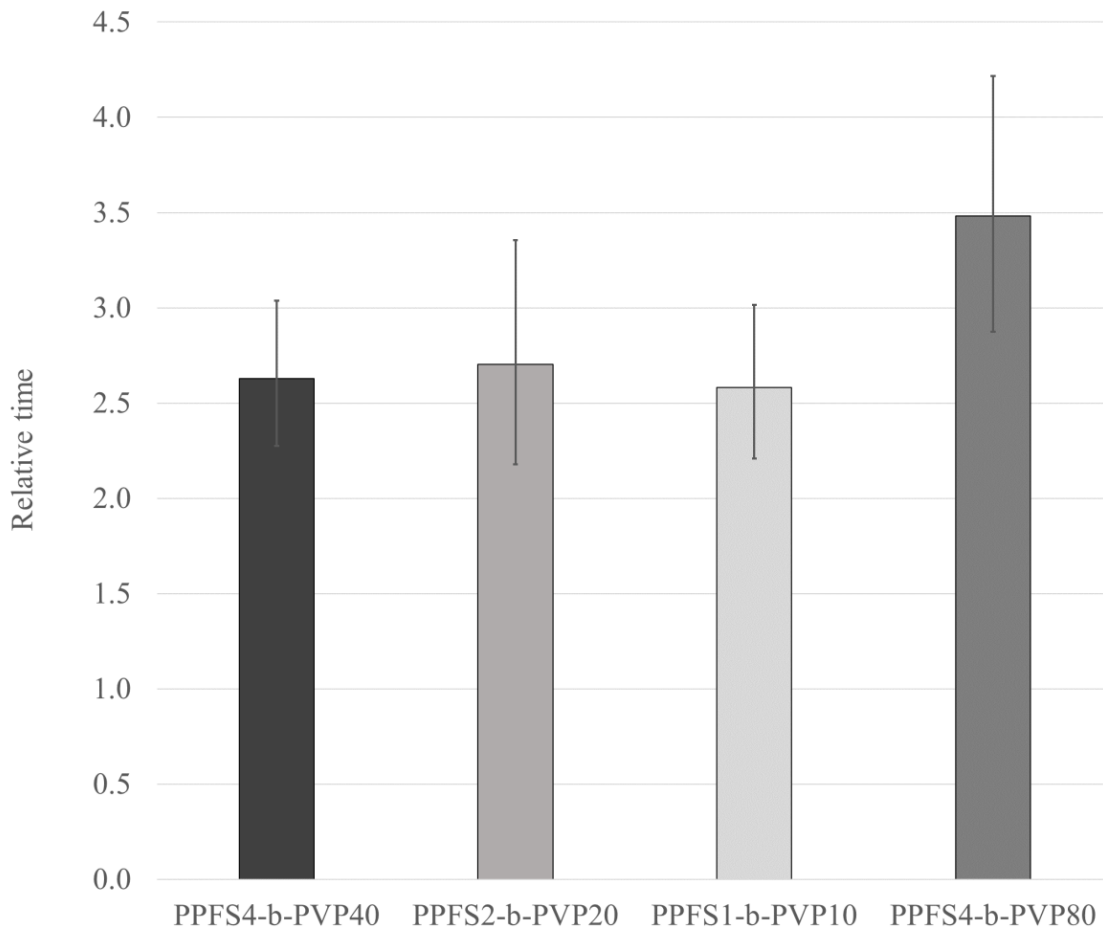

Figure 9 Geometric mean of relative time for methane hydrate slurries to reach to 200 mPa·s for PPFS4-b-PVP, PPFS2-b-PVP20, PPFS1-b-PVP10 and PPFS4-b-PVP80 at 7000 ppm, with 95% confidence interval error bars.

Previous investigations have offered suggestions as to why increasing the concentration 10-fold did not significantly improve the particle dispersion. One suggestion was that the adhesion effects caused by the higher PVP concentration overshadowed the effect of adding a hydrophobic end group, as observed in previous work where agglomerations were the most sudden at 7000 ppm compared to 700 ppm and water systems.[14] Another possible explanation concerns the morphology of the hydrates, which was very different at high block copolymer concentrations and exhibited a smooth, white foamy appearance rather than a porous matrix, possibly due to the formation of soapy bubbles under constant stirring. It is possible that due to the small size of the measuring geometry, the volume expansion due to foam formed at the interface was significant, accelerating the integration of gas molecules and particle agglomeration.

A final possibility is that the high hydrophobic content itself in the system may be unfavorable for water structure disruption. Sa et al. (2013)[18] found that higher-hydrophobicity amino acids with longer alkyl side chain were less effective hydrate inhibitors than the smaller, lower-hydrophobicity amino acids. The former strengthened local water structure while the latter perturbed local water structure. Drawing an analogy to the current study, greater hydrophobic content in water may not always be beneficial to molecular interactions at the water-hydrate interface, and there may be an optimal percentage and chain length for hydrophobic side groups. Reyes et al. (2015)[19] suggested that more significant hydrate inhibition can be achieved through particular spacing between the monomeric units. To verify this hypothesis, block copolymer PPFS4-b-PVP80 was synthesized and tested as a KHI with only 5 wt.% hydrophobic block. As seen in Figure 9, the $t_{200}$ for this less hydrophobic copolymer reached 3.5. This greater value compared to the 10 wt.% PPFS copolymers seemingly contradicts the findings in literature, where increasing hydrophobic block length in water-soluble polymers often resulted in greater inhibition performance. However, the selection of hydrophobic monomers is different in various reports listed in the literature review, and PFS being very hydrophobic might be above a certain optimal threshold at this relatively high concentration.

A final remark is that the presence of aggregates or micelles did not affect the inhibition effect. After 24 hours of stirring, the 7000 ppm solution of PPFS4-b-PVP40 displayed a white, milky color. A shorter copolymer with the same composition, PPFS2-b-PVP20, was synthesized. The solution was still slightly translucent despite the CMC measured by Rajput et al. (2018)[9] being around 10,000 ppm. Although a similar polymerization technique was used, the polymer characterization methods for PPFS2-b-PVP20 were different, their report using the PMMA standards in THF GPC to obtain the $\bar{M}_n$.[9] Finally, the solution with PPFS1-b-PVP10 exhibited a transparent homogeneous appearance at room temperature and was slightly yellow due to the RAFT agent end group. The Z-Average mean measured by DLS for the translucent samples at 40,000 and 20,000 g/mol were in the $10^3$ nm range and were outside of the reliable domain for this equipment. For the clear sample of 10,000 g/mol copolymer, the Z-Average was 43.3 ± 0.1 nm with a PDI of 0.28 ± 0.02. The particle size of the lowest MW copolymer solution was two orders of magnitude lower, but this difference did not significantly affect its inhibitory performance.

The amphiphilic block copolymers poly(ε-caprolactone-b-vinylcaprolactam) by Wan and Liang (2021)[16] were used at concentration of 5,000 ppm, which was above the CMCs, and the 10% hydrophobic block copolymers in micelle form were effective in increasing the nucleation time, roughly by two orders of magnitude compared to pure water and one order of magnitude compared to PVCap. Their molecular weights, however, were around 5,000-6,000 g/mol, deliberately kept low as Colorado School of Mines have demonstrated that PVCap with molecular weights as low as 900 g/mol were more effective than those with higher molecular weights.[4] Additionally, for the same molecular weight of 6,000 g/mol, Seo et al. (2017)[27] measured a much longer nucleation time with PVCap synthesized via RAFT polymerization with low dispersity (Đ < 1.5), while nucleation time for a 6,000 g/mol free-radical polymerized PVCap with high dispersity were quite low and comparable with those of RAFT polymerized PVCap at 10,000 and 20,000 g/mol. The chain lengths studied in this report, 10,000 to 40,000 g/mol, in spite of their dispersity difference, do not affect the KHI's inhibition performance during the growth phase. Posteraro et al. (2015)[8] also came to a similar conclusion in terms of methane consumption rate with commercial PVP from 10,000 to 360,000 g/mol at concentrations from 0.7 to 20,000 ppm.

## 4. Conclusion

Amphiphilic block copolymers PS/PPFS-b-PVP were synthesized using the switchable RAFT agent 2-cyanopropan-2-yl N-methyl-N-(pyridine-4-yl)carbamodithioate and were targeted as KHIs. Chain quality control and low dispersities down to 1.15 and 1.20 for PS and PPFS oligomers, respectively, were achieved with a stoichiometric amount of high purity trifluoromethanesulfonic acid in acetonitrile to protonate the RAFT agent. The dynamic viscosity of methane hydrates formed in the inhibited systems with 700 ppm polymeric additives in reverse-osmosis purified water were recorded for temperatures of 0, 2, 4, 6 °C and pressures of 10 and 15 MPag.

With 10 wt.% of PS or PPFS hydrophobic block, at 700 ppm, the copolymers were able to delay methane hydrate growth by 2.4 and 2.2 times the water reference system, respectively, while the PVP homopolymer at a comparable molecular weight of 40,000 g/mol only attained 1.3 times that of the reference values. There was some evidence that these block copolymers could dissociate hydrate nuclei at low viscosities, though the effect was not reproducible. Increasing the KHI concentration to 7000 ppm only improved the inhibition effect to 2.6-2.7 times that of the reference values, thus the performance at higher concentrations was not desirable compared to PVP in aqueous solutions. By testing a 5 wt.% PPFS-b-PVP at 7000 ppm, however, the relative time reached 3.5 which is greater than for the 10 wt.% block copolymers. These results are different than previous literature observations, where the greater hydrophobicity of KHIs usually provided better inhibition performance. It was suggested that the optimal hydrophobic content for each copolymer could be dependent on the monomer selection. Finally, the presence of aggregates did not impact the inhibition performance of the amphiphilic block copolymers for the molecular weights present in this study, which were between 10,000 and 40,000 g/mol. Low molecular weight oligomers can be tested in the future to verify whether hydrate particle agglomeration is a function of the chain length of polymer additives.

## Acknowledgements

The authors would like to acknowledge the financial support from the Natural Sciences and Engineering Research Council of Canada (NSERC) through Discovery grant number 206269, Discovery grant number 206259, and the Faculty of Engineering of McGill University (MEDA, Vadasz Scholars Program).

**TOC Graphic**

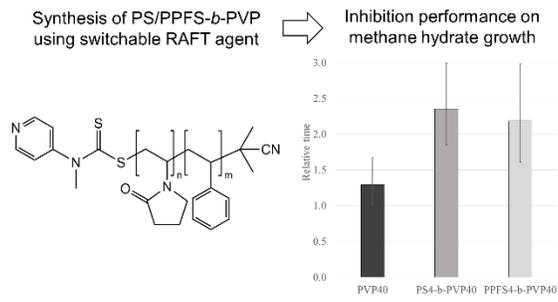